\documentclass[conference]{IEEEtran}
\IEEEoverridecommandlockouts

\newcommand{\IEEEAcceptedNotice}{%
To appear in the Proceedings of the 2026 IEEE Global Communications Conference (GLOBECOM 2026).\\
\copyright~2026 IEEE. Personal use of this material is permitted.  Permission from IEEE must be obtained for all other uses, in any current or future media, including reprinting/republishing this material for advertising or promotional purposes, creating new collective works, for resale or redistribution to servers or lists, or reuse of any copyrighted component of this work in other works.
}

\makeatletter
\def\ps@IEEEtitlepagestyle{%
  \def\@oddfoot{%
    \parbox[b]{0.96\textwidth}{%
      \centering
      \scriptsize
      \IEEEAcceptedNotice
    }%
  }%
  \def\@evenfoot{}%
}
\makeatother

\usepackage{cite}
\usepackage{amsmath,amssymb,amsfonts}
\usepackage{algorithmic}
\usepackage{graphicx}
\usepackage{textcomp}
\usepackage{xcolor}
\usepackage{multirow}
\usepackage{tabu}
\usepackage{makecell}
\usepackage{color}
\usepackage{subfigure}
\usepackage{bm}
\usepackage{balance}

\usepackage[hyphens]{url}
\usepackage{hyperref}
\hypersetup{hidelinks}

\usepackage[ruled,linesnumbered]{algorithm2e}

\SetAlCapNameFnt{\scriptsize}
\SetAlCapFnt{\scriptsize}

\SetCommentSty{mycommfont}

\def\BibTeX{{\rm B\kern-.05em{\sc i\kern-.025em b}\kern-.08em
    T\kern-.1667em\lower.7ex\hbox{E}\kern-.125emX}}
    
\def\BibTeX{{\rm B\kern-.05em{\sc i\kern-.025em b}\kern-.08em T\kern-.1667em\lower.7ex\hbox{E}\kern-.125emX}}

\begin{document}

\title{A Multi-Objective AutoML-based Efficient Intrusion Detection System for EV Charging Networks
}

\author{\IEEEauthorblockN{Li Yang$^*$$^\dagger$}
\IEEEauthorblockA{ 
$^*$Ontario Tech University, Oshawa, Ontario, Canada \\
$^\dagger$Western University, London, Ontario, Canada \\
Emails: li.yang@ontariotechu.ca; lyang339@uwo.ca
} 
}

\maketitle

\begin{abstract}
Electric Vehicle Charging Systems (EVCSs) are increasingly connected with Internet of Things (IoT) devices, which improves charging intelligence but also expands their exposure to cyber-attacks. Intrusion Detection Systems (IDSs) are essential for securing EV charging networks; however, conventional Machine Learning (ML)-based IDSs often rely on manual model design and mainly optimize detection performance without fully considering inference latency and model size. In this paper, a Multi-Objective Automated ML (MOO-AutoML)-based efficient IDS is proposed for EVCS security. The proposed framework uses a lightweight training strategy and a LightGBM-based automated feature selection method to select compact feature subsets based on accumulated feature importance. Then, Non-dominated Sorting Genetic Algorithm III (NSGA-III) jointly optimizes the feature selection threshold and key LightGBM hyperparameters under three objectives: maximizing weighted F1-score, minimizing 99th percentile inference latency ratio, and minimizing model size ratio. Experiments on CICEVSE2024 and CICIDS2017 show that the proposed MOO-AutoML IDS achieves competitive weighted F1-scores, lower P99 inference latency, and smaller model sizes than the compared methods. Overall, the results indicate that the proposed method can support accurate and efficient intrusion detection for EVCS and IoT security under practical deployment constraints.
\end{abstract}
\section{Introduction}
The rapid growth of Electric Vehicles (EVs) and charging infrastructure has transformed EV Charging Systems (EVCSs) into complex cyber-physical systems that connect EV Supply Equipment (EVSE), charging management platforms, mobile applications, cloud servers, and smart grid interfaces \cite{r1}. Through communication protocols such as the Open Charge Point Protocol (OCPP), EVCSs support functions such as remote authentication, load management, and smart charging services \cite{r1}. However, the connectivity that improves charging intelligence and operational convenience also increases the cyber-attack surface of EVCSs \cite{r2}. Attackers may exploit charging station firmware, backend communication channels, or network services to launch various cyber-attacks, such as denial of charging, flooding, data injection, and scanning \cite{r2}. Since EVCSs are connected to transportation and power infrastructures, cyber-attacks may disrupt charging availability, affect grid operations, and lead to financial losses \cite{r3}.

Intrusion Detection Systems (IDSs) are essential for protecting EVCSs and related Internet of Things (IoT) environments \cite{r3}. Unlike preventive mechanisms such as encryption and authentication, IDSs continuously monitor network traffic to detect malicious activities that can bypass preventive defenses \cite{r4}. In EVCS deployment, an IDS can monitor charger-to-backend communication and identify suspicious traffic associated with specific cyber-attacks. Machine Learning (ML) has been widely adopted for IDS development because it can learn complex and nonlinear patterns from large-scale network data and detect diverse attacks accurately \cite{r5}. 

However, conventional ML-based IDSs still face significant limitations in practical deployment. First, many IDS models achieve sub-optimal performance and depend on manual ML model design and tuning, which require substantial domain expertise and repeated trial and error \cite{r6}. Second, many existing studies mainly optimize detection accuracy, while IDS deployment in practical EVCS and IoT environments also requires low latency and memory consumption \cite{r2}. This constraint is especially relevant for charging stations, edge gateways, and IoT devices, which often operate with limited computing resources and memory. Therefore, IDS models for EVCS cybersecurity should be accurate, automated, efficient, and deployment-aware.

Automated ML (AutoML) provides a promising direction that optimizes ML model performance by automating important ML development stages, including Feature Selection (FS), model selection, and Hyper-Parameter Optimization (HPO) \cite{r5}. Nevertheless, traditional AutoML is often formulated as a Single-Objective Optimization (SOO) process that optimizes detection performance at the expense of model complexity, limiting their practical deployment in IoT systems. Multi-Objective Optimization (MOO) addresses this limitation by searching for solutions that balance competing objectives, such as detection performance and model complexity \cite{r4}. 

Therefore, this paper proposes a MOO-AutoML-based efficient IDS for EVCS and IoT systems. The proposed framework first builds an efficient training subset from the original large-scale training data, while reserving the full test set for comprehensive final evaluation. Then, a LightGBM-based \cite{r7} MOO-AutoFS method ranks features by importance and selects feature subsets through an accumulated importance threshold. Finally, Non-dominated Sorting Genetic Algorithm III (NSGA-III) \cite{r8} is used to jointly optimize the feature selection threshold and key LightGBM hyperparameters to balance three crucial objectives: weighted F1-score, 99th Percentile (P99) inference latency ratio, and model size ratio.

To the best of our knowledge, this is the first work that develops a MOO and AutoML-based IDS specifically for EV charging networks and IoT systems by jointly optimizing FS and ML models under detection performance, inference latency, and model compactness objectives. The main contributions of this paper are summarized as follows:
\begin{enumerate}
\item It proposes a novel deployment-aware MOO-AutoML IDS\footnote{Code for this paper is publicly available at: \url{https://github.com/LiYangHart/MOO-NSGA-III-AutoML-based-Intrusion-Detection-System}} for EVCS and IoT security, jointly optimizing detection effectiveness, inference latency, and model size.
\item It introduces a MOO-AutoFS method that converts feature selection into an accumulated feature importance threshold optimization problem.
\item It employs NSGA-III to generate Pareto-optimal LightGBM IDS models, enabling model deployment in EVCS and IoT systems under different constraints.
\item It evaluates the proposed framework on two public benchmark cybersecurity datasets, CICEVSE2024 \cite{r9} and CICIDS2017 \cite{r10}, and compares it with state-of-the-art IDS models.
\end{enumerate}

The remainder of this paper is organized as follows. Section \ref{S2} reviews related work on optimized and automated ML-based IDSs. Section \ref{S3} describes the proposed MOO-AutoML IDS framework in detail. Section \ref{S4} discusses the experimental setup and analyzes results. Section \ref{S5} concludes the paper.

\section{Related Work} \label{S2}

Optimized and automated IDSs have been widely studied to improve cyber-attack detection performance in modern networks and IoT. Elmasry \textit{et al.} \cite{r11} proposed a double Particle Swarm Optimization (PSO)-based method to optimize deep learning models for network intrusion detection. Naeem \textit{et al.} \cite{r12} developed a deep transfer learning and Genetic Algorithm (GA) based framework for multiclass vehicle security. These studies show the value of optimization in IDS design, but their deep learning models can introduce substantial training and computational overhead.

Several works have also explored AutoML for IDSs. Khan \textit{et al.} \cite{r13} proposed an AutoML-based soft voting ensemble model for network intrusion detection. Singh \textit{et al.} \cite{r14} introduced AutoML-ID to automatically select and optimize ML models for wireless sensor network intrusion detection. These methods reduce manual model design and improve detection performance. However, they mainly optimize accuracy and do not explicitly consider inference latency or model size, which are important for EVCS and IoT deployment.

MOO-based IDSs have been developed to balance detection performance and efficiency. Subramani \textit{et al. } \cite{r15} proposed a multi-objective PSO (MOPSO) based feature selection method for IoT-based wireless sensor networks. Yang \textit{et al.} \cite{r16} proposed an intelligent AutoML framework for autonomous intrusion detection. However, existing studies rarely optimize model performance and complexity by considering deployment constraints in EV charging networks.

Therefore, this paper proposes an efficient MOO-AutoML-based IDS for EVCS and IoT security. Unlike existing works, the proposed framework jointly optimizes LightGBM-based AutoFS and model hyperparameters using NSGA-III, while balancing weighted F1-score, P99 inference latency ratio, and model size ratio for deployment-aware intrusion detection.

\section{Proposed MOO-AutoML IDS Framework} \label{S3}
\subsection{System Overview}
The proposed MOO-AutoML IDS is an efficient and accurate IDS that maintains high detection performance while reducing inference latency and model storage cost. This is important for practical deployment in resource-constrained network environments like EVCS and IoT systems, where IDS models must process traffic quickly with limited memory and computing resources. 

\begin{figure}
     \centering
     \includegraphics[width=8cm]{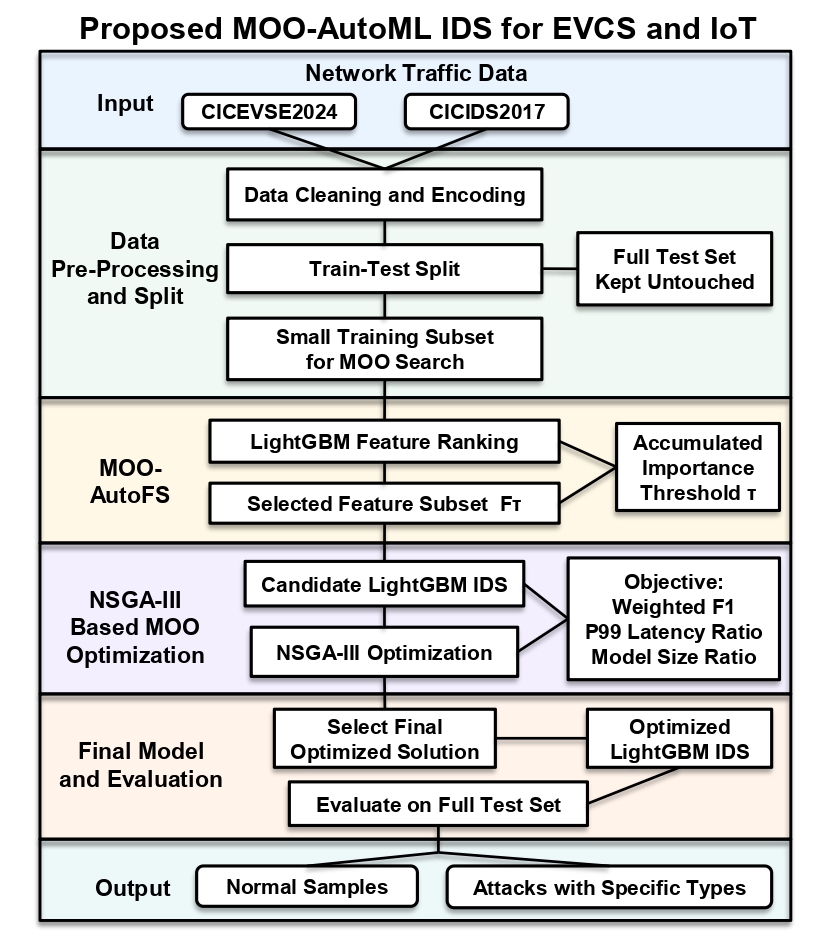}
     \caption{The overview of the proposed MOO-AutoML IDS.} \label{framework}
\end{figure}

As illustrated in Fig. \ref{framework}, the framework takes network traffic data from CICEVSE2024 \cite{r9} and CICIDS2017 \cite{r10} as input. The raw data are first preprocessed and divided into training and test sets. To reduce the computational cost of AutoML search, only a small subset of the training data is used for the MOO process with cross-validation, while the full test set remains untouched for final evaluation. The proposed pipeline then performs LightGBM-based feature ranking, accumulated-importance-based automated feature selection, and NSGA-III-based multi-objective optimization. Each candidate IDS is evaluated using three objectives: weighted F1-score, P99 inference latency ratio, and model size ratio. The final model is selected from the Pareto front and evaluated on the complete test set.

\subsection{Data Pre-Processing and Efficient Training Design}
The raw network traffic datasets include numerical flow features, categorical features, and potentially missing values. Missing numerical values are imputed with feature means, while missing categorical values are imputed using the most frequent category. Categorical or string features are encoded into integer values, as many ML models may fail to process string features directly. Label encoding is chosen because it avoids unnecessary dimensionality increase for the purpose of efficient training \cite{r3}. 

After pre-processing, each dataset is divided into a training set and a test set. A key design in this work is that only a small percentage (\textit{e.g.}, 2\%) of the training set is used during the MOO search to reduce model development cost, while the complete test set is kept unseen until the final evaluation. Since network traffic datasets are usually large and contain many redundant samples, this design enables efficient search while still testing whether the optimized IDS generalizes to a large-scale test distribution.

\subsection{Proposed ML Models and Parameters for Optimization}
After data pre-processing, Feature Selection (FS) is implemented as an important procedure in IDS development, because network traffic datasets often contain irrelevant or redundant features. Reducing these features can reduce inference latency and memory usage, but removing too many features can discard important attack indicators and reduce detection performance. Therefore, the proposed framework introduces a MOO-based Automated FS (AutoFS) method, named MOO AutoFS, to select an effective feature subset automatically.

To enable the proposed AutoFS method, a reference LightGBM model is first trained on the training set to obtain feature importance scores, because LightGBM records the gain contributed by feature splits during tree construction \cite{r17}. These raw importance scores are normalized into relative importance scores, and NSGA-III samples a target accumulated importance threshold $\tau$ to select the smallest top-ranked feature subset whose accumulated importance reaches $\tau$:
\begin{equation}
AI_m=\sum_{i=1}^{m}\frac{I_i}{\sum_{j=1}^{d}I_j}, \quad
S_{\tau}=\{f_1,\ldots,f_m\} ; \text{s.t. } AI_m \geq \tau .
\label{eq:autofs_selected_feature_subset}
\end{equation}
where $I_i$ is the original importance score of feature $i$, $d$ is the total number of features, and $S_{\tau}$ is the selected feature subset from the most important feature $f_1$ to the $m$-th most important feature $f_m$.

This strategy converts feature selection into a compact and interpretable optimization variable. Instead of searching over a binary decision vector for all features, the optimizer only needs to tune the target accumulated importance threshold. A smaller $\tau$ usually selects fewer features and improves efficiency, while a larger $\tau$ retains more predictive information and may improve detection performance. This design makes AutoFS more compact and interpretable, while still linking the selected feature subset to the accumulated predictive contribution of the input features.

After MOO-AutoFS generates a candidate feature subset, a LightGBM classifier is trained as the IDS model for attack detection. LightGBM is selected because it is a highly efficient Gradient Boosting Decision Tree (GBDT) framework that is well-suited for large-scale and high-dimensional tabular data \cite{r7}. Network traffic data are naturally tabular and usually consist of a large number of network features. Tree-based ensemble models, such as LightGBM, can capture nonlinear decision boundaries and feature interactions in such data without requiring the heavy computational cost of deep neural networks \cite{r18}. Moreover, LightGBM uses histogram-based tree learning, Gradient-based One-Side Sampling (GOSS), and Exclusive Feature Bundling (EFB) to reduce training cost and feature dimensionality while preserving informative data points \cite{r7}. These mechanisms help LightGBM achieve strong accuracy while maintaining lower training time and memory usage, making it particularly suitable for the proposed MOO-AutoML framework.

In this work, LightGBM is trained as a multiclass classifier to identify benign traffic and different attack types. The optimized hyperparameters include the number of estimators, learning rate, number of leaves, and maximum depth, because these parameters directly influence model accuracy, inference latency, and model size \cite{r4}. For example, deeper trees and more estimators may increase detection performance but also increase latency and memory usage. Therefore, they should be optimized jointly using advanced optimization methods rather than manual tuning.

The proposed framework formulates IDS model development as a three-objective optimization problem. Each candidate solution consists of a feature selection threshold $\tau$ and a LightGBM hyperparameter configuration $\theta$. The optimizer searches for solutions that maximize detection performance while minimizing deployment cost. The first objective is the weighted F1 score, which is selected because IDS datasets are usually imbalanced and accuracy alone may be misleading when dominant classes account for most samples. F1 score balances precision and recall, and the weighted version further accounts for class support:
\begin{equation}
F1_w=\sum_{c=1}^{C}\frac{n_c}{N}\cdot
\frac{2P_cR_c}{P_c+R_c},
\label{eq:weighted_f1}
\end{equation}
where $C$ is the number of classes, $n_c$ is the number of samples in class $c$, $N$ is the total number of samples, and $P_c$ and $R_c$ are the precision and recall of class $c$. Weighted F1 is used as the main performance objective because it captures both false alarms and missed detections while reflecting the actual class distribution.

The second and third objectives are the P99 inference latency ratio and the model size ratio, respectively:
\begin{equation}
R_{P99}=\frac{L_{P99}^{cand}}{L_{P99}^{base}}, \quad
R_{size}=\frac{S^{cand}}{S^{base}}.
\label{eq:efficiency_ratios}
\end{equation}
where $L_{P99}^{cand}$ and $S^{cand}$ are the candidate model's 99th percentile prediction latency and serialized size, while $L_{P99}^{base}$ and $S^{base}$ are obtained from the full-feature baseline LightGBM model. Many IDS studies report average inference time, but average latency does not fully reflect real-time deployment risk. Therefore, this work uses P99 latency to measure tail inference behavior. A ratio smaller than 1 indicates that the candidate model is more efficient than the baseline. For tree ensemble models, the serialized model size is mainly affected by the number of trees, tree depth, and number of leaves. Thus, minimizing model size encourages compact IDS models that are easier to deploy in edge, IoT, and EVCS environments.

\subsection{NSGA-III based MOO-AutoML Process}
NSGA-III \cite{r8} is used as the MOO engine of the proposed MOO-AutoML IDS because this framework aims to optimize three conflicting objectives: weighted F1-score, P99 inference latency ratio, and model size ratio. Unlike the SOO process that focuses on only one objective or a weighted-sum MOO formulation that combines multiple objectives into a manually weighted score, NSGA-III generates a Pareto front of nondominated solutions, allowing the final IDS model to be selected according to deployment requirements \cite{r4} \cite{r8}.

In the proposed framework, each individual in the NSGA-III population represents one candidate IDS configuration $(\tau,\theta)$, where $\tau$ is the accumulated feature importance threshold used by MOO-AutoFS and $\theta$ is the LightGBM hyperparameter vector. For each candidate, the selected features are determined by $\tau$, a LightGBM model is trained using $\theta$, and the candidate is evaluated using three objectives. Since the weighted F1-score is maximized while latency and model size are minimized, the objective vector is expressed in minimization form as \cite{r8}:
\begin{equation}
\mathbf{g}(\tau,\theta)=
\left[
1-F1_w(\tau,\theta),\;
R_{P99}(\tau,\theta),\;
R_{size}(\tau,\theta)
\right].
\label{eq:nsga3_objective_vector_short}
\end{equation}

The NSGA-III optimization process begins by generating an initial population of candidate IDS configurations. Each candidate is evaluated using cross-validation on the sampled training data. After evaluation, NSGA-III generates offspring solutions through evolutionary search operators and combines the parent and offspring populations. The combined population is then sorted into successive nondominated fronts $F_1, F_2, \ldots$. The first front $F_1$ contains solutions that are not dominated by any other solution. The second front $F_2$ contains solutions dominated only by members of $F_1$, and this process continues until all individuals are ranked. Complete fronts are added to the next population in order of rank until adding another front would exceed the population size. When the last accepted front exceeds the population size, NSGA-III uses reference-point based selection to preserve diverse trade-offs across the three objectives.

For each trial, the optimizer samples $\tau$ and $\theta$, trains the corresponding LightGBM model, evaluates the three objective values, and updates the population. After the predefined number of trials, the nondominated solutions are collected as the Pareto front. The Pareto front is further assessed using hypervolume, which measures the objective space dominated by the nondominated solutions relative to a reference point \cite{r19}. A larger hypervolume indicates that the Pareto front provides stronger overall trade-offs across detection performance, inference latency, and model compactness.

\begin{algorithm}[t]
{\scriptsize
\caption{Proposed NSGA-III Optimized MOO-AutoML IDS}
\label{alg:moo_automl_ids}
\LinesNumbered

\KwIn{
\\\quad $\mathcal{D}=\{X,Y\}$: Network traffic dataset.
\\\quad $\alpha$: class-aware sampling rule that retains all samples from classes with fewer than 10,000 instances and samples 2\% from larger classes.
\\\quad $\Theta$: LightGBM hyperparameter search space.
\\\quad $\mathcal{T}$: accumulated feature importance threshold search space.
\\\quad $T$: number of NSGA-III trials.
\\\quad $K$: number of cross-validation folds.
}

\KwOut{
\\\quad $\mathcal{M}^{*}$: optimized LightGBM IDS.
\\\quad $\mathcal{F}^{*}$: selected feature subset.
\\\quad $\boldsymbol{\theta}^{*}$: optimized hyperparameter configuration.
}

\tcp{Step 1: Data splitting and efficient training subset construction}
Split $\mathcal{D}$ into $\mathcal{D}_{train}^{full}$ and $\mathcal{D}_{test}$\;
Sample $\mathcal{D}_{train}^{sub}$ from $\mathcal{D}_{train}^{full}$ using ratio $\alpha$\;
Preprocess $\mathcal{D}_{train}^{sub}$ and $\mathcal{D}_{test}$\;

\tcp{Step 2: Feature importance estimation}
Train a reference LightGBM model on $\mathcal{D}_{train}^{sub}$\;
Compute feature importance scores $\{I_i\}_{i=1}^{d}$\;
$RI_i \leftarrow \frac{I_i}{\sum_{j=1}^{d}I_j}$ for each feature $i$\;
Sort features according to $RI_i$ in descending order\;
Compute accumulated importance scores $\{AI_m\}_{m=1}^{d}$\;

\tcp{Step 3: Baseline model for normalized efficiency objectives}
Train a full-feature baseline LightGBM model on $\mathcal{D}_{train}^{sub}$\;
Record baseline P99 latency $L_{P99}^{base}$ and baseline model size $S^{base}$\;

\tcp{Step 4: NSGA-III based MOO-AutoFS and LightGBM optimization}
Initialize the NSGA-III optimizer\;
\For{$t\leftarrow 1$ \KwTo $T$}{
    Sample $\tau_t \in \mathcal{T}$ and $\boldsymbol{\theta}_t \in \Theta$\;
    Select $\mathcal{F}_{\tau_t}$ such that $AI_m \geq \tau_t$\;
    Train LightGBM with $\mathcal{F}_{\tau_t}$ and $\boldsymbol{\theta}_t$ using $K$-fold cross-validation\;
    Compute weighted F1-score $F1_w$\;
    Measure candidate P99 latency $L_{P99}^{cand}$ and compute $R_{P99}=\frac{L_{P99}^{cand}}{L_{P99}^{base}}$\;
    Compute candidate model size $S^{cand}$ and model size ratio $R_{size}=\frac{S^{cand}}{S^{base}}$\;
    Return objective vector $[1-F1_w, R_{P99}, R_{size}]$ to NSGA-III\;
}

\tcp{Step 5: Extract final Pareto solution}
Extract the nondominated Pareto front from all trials\;
Compute hypervolume to evaluate Pareto front quality\;
Select the final deployment-oriented Pareto solution $(\mathcal{F}^{*},\boldsymbol{\theta}^{*})$\;

\tcp{Step 6: Final model training and evaluation}
Train $\mathcal{M}^{*}$ on $\mathcal{D}_{train}^{sub}$ using $\mathcal{F}^{*}$ and $\boldsymbol{\theta}^{*}$\;
Evaluate $\mathcal{M}^{*}$ on the complete test set $\mathcal{D}_{test}$\;
\Return $\mathcal{M}^{*}$, $\mathcal{F}^{*}$, and $\boldsymbol{\theta}^{*}$\;

}
\end{algorithm}

Overall, the proposed MOO-AutoML IDS has several advantages:
\begin{enumerate}
\item It explicitly optimizes both detection effectiveness and deployment efficiency, rather than treating latency and model size as secondary metrics after model training. 
\item The proposed MOO AutoFS method provides an efficient and interpretable feature selection process based on accumulated relative feature importance. 
\item The P99 latency is used to capture tail inference behavior, which is more important than average latency for real-time IDS deployment. 
\item Model size ratio directly reflects memory and storage efficiency. 
\item Using only a small percentage (\textit{i.e.}, 2\%) of the training set substantially reduces AutoML search cost, while evaluation on the full test set provides comprehensive validation. 
\item The Pareto front provides multiple deployment choices, allowing the final model to be selected according to different operational requirements.
\end{enumerate}

\section{Performance Evaluation} \label{S4}
\subsection{Experimental Setup}
The proposed IDS was developed and evaluated by extending the Scikit-learn, LightGBM \cite{r7}, and Optuna \cite{r20} libraries in Python 3.7. The experiments were conducted on an Alienware Aurora R9 machine with an Intel Core i9-9900K processor, 64 GB RAM, and an RTX 2080 Ti GPU, representing a central server machine in EVCSs or IoT systems.

The proposed framework is evaluated on two public cybersecurity datasets, CICEVSE2024 \cite{r9} and CICIDS2017 \cite{r10}. CICEVSE2024 \cite{r9} is designed for EV charging station cybersecurity research and contains benign traffic and various attack scenarios collected from EVCSs under both idle and charging states, including 12 specific types of DoS, reconnaissance, and scanning attacks. CICIDS2017 \cite{r10} is a widely used IDS benchmark dataset that contains benign traffic and many general modern attacks, including DoS, bot, brute-force, port-scan, infiltration, and web-attacks \cite{r2}. Both datasets are large-scale network traffic datasets, each containing more than two million samples. These two datasets are selected to evaluate EVCS-specific security performance and generalizability to broader network attack detection.

For both original large datasets, an 80/20\% train-test split is used for hold-out validation, and only 2\% of the original training set is used as the final training set for efficient model learning and optimization, while the complete large-scale test set is used for final evaluation. For CICIDS2017, the sampled training set contains 48,569 samples with 77 features, and the full test set contains 566,149 samples. For CICEVSE2024, the sampled training set contains 47,338 samples with 85 features, and the full test set contains 548,940 samples. The MOO process uses three-fold cross-validation on the sampled training set. NSGA-III is implemented through Optuna with a total budget of 50 trials and a fixed random seed of 42. The model performance metrics include accuracy, precision, recall, and weighted F1-score. Weighted F1 is used as the main performance metric because the datasets are imbalanced. Efficiency metrics include average inference latency per sample, P99 inference latency, and model size. 

\subsection{Experimental Results and Discussion}

\begin{table}[!t]
\caption{Performance Evaluation of The Models on CICEVSE2024}
\centering
\setlength\extrarowheight{1pt}
\scalebox{0.685}{
\begin{tabular}{|>{\centering\arraybackslash}p{7em}|
>{\centering\arraybackslash}p{3.7em}|
>{\centering\arraybackslash}p{3.7em}|
>{\centering\arraybackslash}p{3.2em}|
>{\centering\arraybackslash}p{3.8em}|
>{\centering\arraybackslash}p{4.1em}|
>{\centering\arraybackslash}p{4.1em}|
>{\centering\arraybackslash}p{3.0em}|}
\hline
\textbf{Method} & \textbf{Accuracy (\%)} & \textbf{Precision (\%)} & \textbf{Recall (\%)} & \textbf{Weighted F1 (\%)} & \textbf{Avg Inference Latency Per Sample (ms)} & \textbf{P99 Inference Latency (ms)} & \textbf{Model Size (MB)} \\
\hline
PSO-LSTM \cite{r11} & 88.849 & 91.093 & 88.849 & 87.486 & 0.0175 & 0.0170 & 4.83 \\
\hline
GA-CNN \cite{r12} & 96.605 & 96.622 & 96.605 & 96.560 & 0.0133 & 0.0143 & 4.59 \\
\hline
OE-IDS \cite{r13} & 99.936 & 99.954 & 99.936 & 99.944 & 0.0626 & 0.0702 & 33.40 \\
\hline
AutoML-ID \cite{r14} & 99.917 & 99.925 & 99.917 & 99.920 & 0.0407 & 0.0512 & 3.14 \\
\hline
MOPSO-SVM \cite{r15} & 98.322 & 98.380 & 98.322 & 98.292 & 0.0562 & 0.0576 & 8.38 \\
\hline
Baseline LightGBM \cite{r7} & 99.767 & 99.777 & 99.767 & 99.771 & 0.0132 & 0.0156 & 6.07 \\
\hline
SOO-AutoML \cite{r16} & 99.936 & 99.954 & 99.936 & 99.944 & 0.0118 & 0.0231 & 4.05 \\
\hline
Proposed MOO-AutoML & 99.935 & 99.940 & 99.936 & 99.937 & 0.0107 & 0.0112 & 2.52 \\
\hline
\end{tabular}
}
\label{tab:cicevse2024_results}
\end{table}

\begin{table}[!t]
\caption{Performance Evaluation of The Models on CICIDS2017}
\centering
\setlength\extrarowheight{1pt}
\scalebox{0.685}{
\begin{tabular}{|>{\centering\arraybackslash}p{7em}|
>{\centering\arraybackslash}p{3.7em}|
>{\centering\arraybackslash}p{3.7em}|
>{\centering\arraybackslash}p{3.2em}|
>{\centering\arraybackslash}p{3.8em}|
>{\centering\arraybackslash}p{4.1em}|
>{\centering\arraybackslash}p{4.1em}|
>{\centering\arraybackslash}p{3.0em}|}
\hline
\textbf{Method} & \textbf{Accuracy (\%)} & \textbf{Precision (\%)} & \textbf{Recall (\%)} & \textbf{Weighted F1 (\%)} & \textbf{Avg Inference Latency Per Sample (ms)} & \textbf{P99 Inference Latency (ms)} & \textbf{Model Size (MB)} \\
\hline
PSO-LSTM \cite{r11} & 98.142 & 98.742 & 98.142 & 98.407 & 0.0203 & 0.0361 & 1.46 \\
\hline
GA-CNN \cite{r12} & 99.596 & 99.679 & 99.596 & 99.625 & 0.0147 & 0.0138 & 1.93 \\
\hline
OE-IDS \cite{r13} & 99.693 & 99.714 & 99.693 & 99.698 & 0.0255 & 0.0380 & 52.25 \\
\hline
AutoML-ID \cite{r14} & 99.330 & 99.413 & 99.330 & 99.359 & 0.0172 & 0.0294 & 0.77 \\
\hline
MOPSO-SVM \cite{r15} & 95.624 & 96.373 & 95.624 & 95.867 & 0.0343 & 0.0553 & 8.28 \\
\hline
Baseline LightGBM \cite{r7} & 99.548 & 99.571 & 99.548 & 99.554 & 0.0040 & 0.0065 & 2.82 \\
\hline
SOO-AutoML \cite{r16} & 99.805 & 99.836 & 99.805 & 99.816 & 0.0087 & 0.0128 & 2.36 \\
\hline
Proposed MOO-AutoML & 99.697 & 99.717 & 99.697 & 99.704 & 0.0030 & 0.0033 & 0.70 \\
\hline
\end{tabular}
}
\label{tab:cicids2017_results}
\end{table}

The experimental results of the proposed MOO-AutoML IDS and the compared state-of-the-art methods \cite{r11}, \cite{r12}, \cite{r13}, \cite{r14}, \cite{r15}, \cite{r16} on the CICEVSE2024 and CICIDS2017 datasets are shown in Tables \ref{tab:cicevse2024_results} and \ref{tab:cicids2017_results}, respectively. The compared methods include optimized deep learning models, AutoML-based IDSs, MOO-based feature selection, the baseline LightGBM model, and the SOO-AutoML method, which optimizes F1-score while not explicitly optimizing model complexity. As shown in the two tables, the proposed framework achieves a strong balance between detection performance and deployment efficiency.

For CICEVSE2024, the proposed MOO-AutoML IDS achieves a high 99.937\% weighted F1-score. It is lower by only 0.007\% weighted F1-score compared with the best-performing SOO-AutoML and OE-IDS models, but achieves the best efficiency among all compared methods. Specifically, the proposed model obtains the lowest average inference latency of 0.0107 ms, the lowest P99 inference latency of 0.0112 ms, and the smallest model size of 2.52 MB. Compared with the baseline LightGBM model, it improves the weighted F1-score from 99.771\% to 99.937\%, while reducing the P99 inference latency from 0.0156 ms to 0.0112 ms and the model size from 6.07 MB to 2.52 MB. This demonstrates that the proposed MOO-AutoFS and NSGA-III optimization can generate a more accurate and compact IDS model for EVCS security. The final optimized LightGBM model selects only 10 features. The optimized hyperparameters are $n\_estimators=150$, $learning\_rate=0.0376$, $num\_leaves=15$, and $max\_depth=6$. This compact feature subset and shallow tree structure explain why the proposed model maintains high detection performance while achieving low latency and small model size.

For CICIDS2017, the proposed MOO-AutoML IDS achieves a high weighted F1-score of 99.704\%. Although the SOO-AutoML model obtains the highest weighted F1-score of 99.816\%, it has a larger P99 inference latency of 0.0128 ms and a model size of 2.36 MB. In contrast, the proposed model achieves the lowest average inference latency of 0.0030 ms, the lowest P99 inference latency of 0.0033 ms, and the smallest model size of 0.70 MB. Compared with the baseline LightGBM model, it improves the weighted F1-score from 99.554\% to 99.704\%, while reducing the P99 inference latency from 0.0065 ms to 0.0033 ms and the model size from 2.82 MB to 0.70 MB. The proposed model selects 33 features, and the final LightGBM parameters are $n\_estimators=60$, $learning\_rate=0.0493$, $num\_leaves=16$, and $max\_depth=7$. These results show that the proposed framework can adapt the selected feature subset and model complexity according to different dataset characteristics.

Overall, instead of simply maximizing the detection performance, the proposed MOO-AutoML IDS searches for Pareto-optimal models that jointly consider detection effectiveness, tail inference latency, and model compactness. The experimental results indicate that the proposed framework achieves near state-of-the-art detection performance while substantially reducing inference latency and model size, making it suitable for resource-constrained EVCS and IoT environments.

\section{Conclusion} \label{S5}

This paper proposed a MOO-AutoML-based efficient IDS for EV charging network security. The proposed framework integrates lightweight training, LightGBM-based automated feature selection, and NSGA-III-based multi-objective optimization to jointly optimize detection performance, P99 inference latency, and model size. Unlike conventional IDSs that mainly focus on accuracy, the proposed method explicitly considers deployment constraints for EVCS and IoT environments. The proposed framework was evaluated on two public benchmark cybersecurity datasets, CICEVSE2024 and CICIDS2017. Experimental results show that the proposed MOO-AutoML IDS achieves strong weighted F1-scores of 99.937\% and 99.704\%, low P99 inference latencies of 0.0112 ms and 0.0033 ms per sample, and compact model sizes of 2.52 MB and 0.70 MB on the two datasets, respectively. These results demonstrate that the proposed framework can provide accurate, efficient, and deployment-aware intrusion detection for EV charging networks and broader IoT systems. Future work will explore Large Language Model (LLM)-assisted AutoML to further automate model optimization and adapt to evolving deployment requirements.

\section*{Acknowledgment}
This project was made possible in part through the support of the National Cybersecurity Consortium and the Government of Canada (CSIN).


\begin{thebibliography}{00}
\bibitem{r1} Z. Garofalaki, D. Kosmanos, S. Moschoyiannis, D. Kallergis, and C. Douligeris, “Electric Vehicle Charging: A Survey on the Security Issues and Challenges of the Open Charge Point Protocol (OCPP),” \textit{IEEE Commun. Surveys Tuts.}, vol. 24, no. 3, pp. 1504--1533, Sep. 2022.

\bibitem{r2} F. Dehrouyeh, L. Yang, F. Badrkhani Ajaei, and A. Shami, “On TinyML and Cybersecurity: Electric Vehicle Charging Infrastructure Use Case,” \textit{IEEE Access}, vol. 12, pp. 108703--108730, 2024.

\bibitem{r3} L. Yang and G. Kirubavathi, “EGGA: An Error-Guided Generative Augmentation and Optimized ML-Based IDS for EV Charging Network Security,” \textit{Future Internet}, vol. 18, no. 4, Art. no. 202, Apr. 2026.

\bibitem{r4} L. Yang and A. Shami, “Toward Autonomous and Efficient Cybersecurity: A Multi-Objective AutoML-Based Intrusion Detection System,” \textit{IEEE Trans. Mach. Learn. Commun. Netw.}, vol. 3, pp. 1244--1264, Nov. 2025.

\bibitem{r5} L. Yang, M. El Rajab, A. Shami, and S. Muhaidat, “Enabling AutoML for Zero-Touch Network Security: Use-Case Driven Analysis,” \textit{IEEE Trans. Netw. Service Manag.}, vol. 21, no. 3, pp. 3555--3582, 2024.

\bibitem{r6} L. Yang, S. Naser, A. Shami, S. Muhaidat, L. Ong, and M. Debbah, “Toward Zero Touch Networks: Cross-Layer Automated Security Solutions for 6G Wireless Networks,” \textit{IEEE Trans. Commun.}, vol. 73, no. 9, pp. 7650--7679, 2025.

\bibitem{r7} G. Ke, Q. Meng, T. Finley, T. Wang, W. Chen, W. Ma, Q. Ye, and T.-Y. Liu, “LightGBM: A Highly Efficient Gradient Boosting Decision Tree,” \textit{Adv. Neural Inf. Process. Syst.}, vol. 30, pp. 3146--3154, 2017.

\bibitem{r8} K. Deb and H. Jain, “An Evolutionary Many-Objective Optimization Algorithm Using Reference-Point-Based Nondominated Sorting Approach, Part I: Solving Problems With Box Constraints,” \textit{IEEE Trans. Evol. Comput.}, vol. 18, no. 4, pp. 577--601, 2014.

\bibitem{r9} E. D. Buedi, A. A. Ghorbani, S. Dadkhah, and R. L. Ferreira, “Enhancing EV Charging Station Security Using a Multi-dimensional Dataset: CICEVSE2024,” in \textit{Data and Applications Security and Privacy XXXVIII, Lecture Notes in Comput. Sci.}, vol. 14901, 2024, pp. 171--190.

\bibitem{r10} I. Sharafaldin, A. H. Lashkari, and A. A. Ghorbani, “Toward Generating a New Intrusion Detection Dataset and Intrusion Traffic Characterization,” in \textit{Proc. 4th Int. Conf. Inf. Syst. Security Privacy (ICISSP)}, 2018, pp. 108--116.

\bibitem{r11} W. Elmasry, A. Akbulut, and A. H. Zaim, “Evolving Deep Learning Architectures for Network Intrusion Detection Using a Double PSO Metaheuristic,” \textit{Comput. Netw.}, vol. 168, Art. no. 107042, Feb. 2020.

\bibitem{r12} H. Naeem, F. Ullah, O. Krejcar, D. Li, and D. Vasan, “Optimizing Vehicle Security: A Multiclassification Framework Using Deep Transfer Learning and Metaheuristic-Based Genetic Algorithm Optimization,” \textit{Int. J. Crit. Infrastruct. Prot.}, vol. 49, Art. no. 100745, Jul. 2025.

\bibitem{r13} M. A. Khan, N. Iqbal, Imran, H. Jamil, and D. H. Kim, “An Optimized Ensemble Prediction Model Using AutoML Based on Soft Voting Classifier for Network Intrusion Detection,” \textit{J. Netw. Comput. Appl.}, vol. 212, Art. no. 103560, Mar. 2023.

\bibitem{r14} A. Singh, J. Amutha, J. Nagar, S. Sharma, and C.-C. Lee, “AutoML-ID: Automated Machine Learning Model for Intrusion Detection Using Wireless Sensor Network,” \textit{Sci. Rep.}, vol. 12, no. 1, Art. no. 9074, May 2022.

\bibitem{r15} S. Subramani and M. Selvi, “Multi-Objective PSO Based Feature Selection for Intrusion Detection in IoT Based Wireless Sensor Networks,” \textit{Optik}, vol. 273, Art. no. 170419, Feb. 2023.

\bibitem{r16} L. Yang and A. Shami, “Towards Autonomous Cybersecurity: An Intelligent AutoML Framework for Autonomous Intrusion Detection,” in \textit{Proc. Workshop Autonomous Cybersecurity (AutonomousCyber ’24), ACM SIGSAC Conf. Comput. Commun. Security (CCS ’24)}, 2024, pp. 1--11.

\bibitem{r17} L. Yang, A. Shami, G. Stevens, and S. DeRusett, “LCCDE: A Decision-Based Ensemble Framework for Intrusion Detection in the Internet of Vehicles,” in \textit{Proc. 2022 IEEE Glob. Commun. Conf. (GLOBECOM)}, 2022, pp. 1--6.

\bibitem{r18} L. Grinsztajn, E. Oyallon, and G. Varoquaux, “Why Do Tree-Based Models Still Outperform Deep Learning on Typical Tabular Data?,” in \textit{Adv. Neural Inf. Process. Syst.}, vol. 35, 2022, pp. 507--520.

\bibitem{r19} E. Zitzler and L. Thiele, “Multiobjective Evolutionary Algorithms: A Comparative Case Study and the Strength Pareto Approach,” \textit{IEEE Trans. Evol. Comput.}, vol. 3, no. 4, pp. 257--271, Nov. 1999.

\bibitem{r20} T. Akiba, S. Sano, T. Yanase, T. Ohta, and M. Koyama, “Optuna: A Next-Generation Hyperparameter Optimization Framework,” in \textit{Proc. 25th ACM SIGKDD Int. Conf. Knowl. Discovery Data Mining}, 2019, pp. 2623--2631.


\end{thebibliography}
\end{document}